\documentstyle[12pt]{article}

\makeatletter

\newcommand {\beq}{\begin{equation}}
\newcommand {\eeq}{\end{equation}}
\newcommand {\beqa}{\begin{eqnarray}}
\newcommand {\eeqa}{\end{eqnarray}}         
\newcommand {\beqs}{\begin{eqnarray*}}
\newcommand {\eeqs}{\end{eqnarray*}}
\newcommand {\bds}{\begin{displaymath}}
\newcommand {\eds}{\end{displaymath}}
\newcommand {\n}{\nonumber\\}

\newcommand {\bebb}{\begin{thebibliography}}
\newcommand {\eebb}{\end{thebibliography}}      
\newcommand {\bbit}{\bibitem}

\begin{document}


\begin{flushright}
\end{flushright}

\vskip 1cm

\begin{center}
{\large\bf On analytic solutions of the driven, 2-photon and two-mode quantum Rabi models} 

\vspace{1cm}

{Yao-Zhong Zhang}$\,$\footnote{Email: yzz@maths.uq.edu.au}
\vskip.1in

{\em School of Mathematics and Physics,
The University of Queensland\\ Brisbane, Qld 4072, Australia}

\end{center}

\date{}



\begin{abstract}
By applying the Bogoliubov transformations and through the introduction of the Bargmann-Hilbert spaces,
we obtain analytic representations of solutions to the driven Rabi model without ${\cal Z}_2$ symmetry, and the 2-photon and two-mode quantum Rabi models. 
In each case, transcendental function is analytically derived whose zeros give the energy spectrum of the model. The zeros can be numerically found by standard root-search techniques.  
\end{abstract}

\vskip.1in

{\it PACS numbers}:  03.65.Ge, 02.30.Ik, 42.50.Pq

{\it Keywords}: Quasi-exactly solvable systems, Rabi model, Solutions of wave equations










\section{Introduction}
Spin-boson systems describe interactions between spins and harmonic oscillators (boson modes)
and have played a prominent role in modeling the ubiquitous matter-light interactions in modern physics.
One of the well-known spin-boson systems is the quantum Rabi model.  This model describes the interaction
of a two-level atom with a harmonic mode of quantized electromagnetic field. 
Due to the simplicity of its Hamiltonian, the Rabi model has served as the basis
for understanding matter-light interactions, and has a variety of applications ranging from  
quantum optics \cite{Vedral06} to solid state semiconductor systems \cite{Khitrova06} 
and molecular physics \cite{Thanopulos04}. It has also played a significant role in
the novel research field of cavity and circuit quantum electrodynamics \cite{Englund07,Niemczyk10}. 

The Hamiltonian of the Rabi model with ${\cal Z}_2$ symmetry is given by
\beq
H_R=\omega b^\dagger b+\Delta\,\sigma_z+g\,\sigma_x(b^\dagger+b),
\eeq
where $g$ is the spin-boson interaction strength, 
$\sigma_z, \sigma_x$ are the Pauli matrices describing the two atomic levels separated 
by energy difference $2\Delta$, and $b^\dagger$ ($b$) are 
creation (annihilation) operators of a boson mode with frequency $\omega$. 
In the Bargmann space with the monomials $\left\{\frac{w^n}{\sqrt{n!}}\right\}$ as basis 
vectors, the boson creation and annihilation operators can be realized as 
$b^\dagger = w$, $b=\frac{d}{dw}$ \cite{Bargmann61}.
In terms of the two-component wavefunction $\psi(w)=\left(\psi_+(w), \psi_-(w)\right)^T$,
the time-independent Schr\"odinger equation of the model  
gives a system of two coupled linear differential equations with rational coefficients. 
The system depends on a spectral parameter $E$. The wavefunction components are
elements of Bargmann-Hilbert (BH) space of entire functions of one variable $w\in {\bf C}$ \cite{Schweber67}.
The scalar product of any two elements $f(w), g(w)$ in the BH space is given by
\beq
(f,g)=\int\,\overline{f(w)}\,g(w)\,d\mu(w),\label{inner-product1}
\eeq
where $\overline{f(w)}$ is the complex conjugate of $f(w)$ and $d\mu(w)=\frac{1}{\pi}e^{-|w|^2}\,dx\,dy$ is the measure. 
The energy $E$ belongs to the spectrum
of the problem if and only if for this value of $E$ the system of coupled differential equations
(given by the time-independent Schr\"odinger equation) has entire solution \cite{Schweber67}.

Using the BH space approach, Braak \cite{Braak11} recently presented a pair of 
transcendental functions defined as infinite 
power series expansions with coefficients satisfying three-term recurrence relations, 
and argued that the spectrum of the Rabi model is given by the zeros of the transcendental functions. 
This theoretical progress has renewed the interest in the Rabi and related models
\cite{Solano11}-\cite{Zhang17}.

In this work, we apply the Bogoliubov transformations and the BH approach to obtain analytic solutions of the driven Rabi model with broken ${\cal Z}_2$ symmetry and the 2-photon and two-mode quantum Rabi models.
The driven Rabi model is a generalization of the ${\cal Z}_2$ symmetrical Rabi model \cite{Braak11}. It has recently been used in \cite{Larson13} to examine
quantum thermalization. The 2-photon and two-mode Rabi models are phenomenological models describing a two-level atom interacting with 2 photons and 2 harmonic modes,
respectively. They can be experimentally realized in circuit quantum electromagnetic systems \cite{Niemczyk10}, and have established applications in many research fields,
including Rubidium atoms \cite{Brune87} and quantum dots \cite{Valle10,Ota11}. 

Previous studies mainly concerned the Rabi 
model and its simple variations and presumed the standard measure (given above) for the corresponding 
BH spaces. However as seen in sections \ref{2-photon} and \ref{two-mode} below, 
for the 2-photon and two-mode Rabi models, 
the standard  measure is no longer appropriate for defining the scalar products of their BH spaces. This is because the scalar product defined with the standard measure is not
consistent with the differential representations (\ref{2-photon-Rabi-diff}) and 
(\ref{su11-diff-rep-2mode}). In this work, we introduce new BH spaces for the two models and find the suitable measures needed to define their scalar products. This leads to the completely new criteria for
entire (wave)functions, as can be seen from (\ref{entireness-test1}) and (\ref{entireness-test2}) 
[cf. (\ref{entireness-test-for-Rabi}) related to the standard measure]. These criteria 
are essential in the derivation of the analytic solutions presented here for the 2-photon and two-mode Rabi models.

\section{Driven quantum Rabi model}

The Hamiltonian of the driven quantum Rabi model is given by
\beq
H_{dR}=\omega b^\dagger b+\Delta\,\sigma_z+g\,\sigma_x(b^\dagger+b)+\delta\,\sigma_x,\label{RabiH1}
\eeq
where $\delta$ is the drive amplitude. 
After a canonical Bogoliubov transformation $b=a+\frac{\lambda}{\omega},~~b^\dagger=a^\dagger+\frac{\lambda}{\omega}$, 
where $\lambda$ is a real parameter, the Hamiltonian (\ref{RabiH1}) becomes
\beq
\tilde{H}_{dR}=\omega a^\dagger a+\Delta\,\sigma_z+(g\,\sigma_x+\lambda) (a^\dagger+a)
   +\delta\,\sigma_x+\frac{2g\lambda}{\omega}\,\sigma_x+\frac{\lambda^2}{\omega}.\label{Hdr}
\eeq

Using the Bargmann realization $a^\dagger=z,~~a=\frac{d}{dz}$, 
working in a representation defined by $\sigma_x$ diagonal and choosing $\lambda=-g$,
we can turn the transformed driven Rabi Hamiltonian into the matrix differential operator
\beq
\tilde{H}_{dR}=\left(
\begin{array}{cc}
\omega\,z\frac{d}{dz}+\delta-\frac{g^2}{\omega} & \Delta \\
\Delta & \omega\left(z-\frac{2g}{\omega}\right)\frac{d}{dz}-2gz-\delta +\frac{3g^2}{\omega}
\end{array}
\right ).
\eeq
The BH space is the space of entire functions $f(z)$ with inner product
(\ref{inner-product1}) and orthonormal basis $\left\{\frac{z^n}{\sqrt{n!}}\right\}$. So if
$f(z)=\sum_{n=0}^\infty c_n\,z^n$, then 
\beq
||f||^2=\sum_{n=0}^\infty|c_n|^2\,n!\label{entireness-test-for-Rabi}
\eeq
and $f(z)$ is entire analytic function iff this sum converges \cite{Bargmann61}.

In terms of two-component wavefunction $\varphi(z)=(\varphi_+(z), \varphi_-(z))^T$, the
time-independent Schr\"odinger equation, $\tilde{H}_{dR}\,\varphi(z)=E\,\varphi(z)$, yields a system
of couple of differential equations
\beqa
&&\left(\omega\,z\frac{d}{dz}+\delta-\frac{g^2}{\omega}-E\right)\varphi_+ +\Delta\,\varphi_- =0,\label{Rabi-diff1}\\
&&\left[\omega\left(z-\frac{2g}{\omega}\right)\frac{d}{dz}-2gz+\frac{3g^2}{\omega}-\delta-E\right]\varphi_-
    +\Delta\,\varphi_+ =0.\label{Rabi-diff2}
\eeqa
Solutions to these equations must be analytic in the whole complex plane 
if $E$ belongs to the spectrum of $\tilde{H}_{dR}$. So we are looking for
solutions of the form
\beqa
\varphi_+(z)=\sum_{n=0}^\infty {\cal R}_n^+(E)\,z^n,~~~~~~
\varphi_-(z)=\sum_{n=0}^\infty {\cal R}_n^-(E)\,z^n,\label{Rabi-series-solution}
\eeqa
which converge in the entire complex plane and are elements of the BH space with inner product (\ref{inner-product1}).

Substituting (\ref{Rabi-series-solution}) into (\ref{Rabi-diff1}), we obtain
\beq
{\cal R}_n^+=\frac{\Delta}{E-n\omega-\delta+\frac{g^2}{\omega}}\,{\cal R}_n^-.
\eeq
Thus ${\cal R}_n^+$ is not analytic in $E$ but has simple poles at
\beq
E=n\omega+\delta-\frac{g^2}{\omega},~~~~n=0,1,\cdots.
   \label{Rabi-exceptional1}
\eeq
This gives one set of exact energies of the driven Rabi model.

Choosing $\lambda=-g$ in (\ref{Hdr}), we get
\beq
{\tilde{H}}_{dR}=\left(
\begin{array}{cc}
\omega\left(z+\frac{2g}{\omega}\right)\frac{d}{dz}+2gz+\delta +\frac{3g^2}{\omega} & \Delta \\
\Delta & \omega\,z\frac{d}{dz}-\delta-\frac{g^2}{\omega}
\end{array}
\right ).
\eeq
Let  $\phi(z)=(\phi_+(z), \phi_-(z))^T$ be two-component wavefunction. Then the
time-independent Schr\"odinger equation in this case yields 
\beqa
&&\left[\omega\left(z+\frac{2g}{\omega}\right)\frac{d}{dz}+2gz+\frac{3g^2}{\omega}+\delta-E\right]\phi_+    +\Delta\,\phi_- =0.\label{Rabi-diff1b}\\
&&\left(\omega\,z\frac{d}{dz}-\delta-\frac{g^2}{\omega}-E\right)\phi_- +\Delta\,\varphi_+ =0,\label{Rabi-diff2b}.
\eeqa
Similar to the $\lambda=g$ case, solutions to these equations must be analytic in the whole complex plane 
if $E$ belongs to the spectrum of $\tilde{H}_{dR}$. So we search for
solutions of the form
\beqa
\phi_+(z)=\sum_{n=0}^\infty {\cal S}_n^+(E)\,z^n,~~~~~~
\phi_-(z)=\sum_{n=0}^\infty {\cal S}_n^-(E)\,z^n,\label{Rabi-series-solution2}
\eeqa
which are elements of the BH space.

Substituting (\ref{Rabi-series-solution2}) into (\ref{Rabi-diff2b}), we obtain
\beq
{\cal S}_n^-=\frac{\Delta}{E-n\omega+\delta+\frac{g^2}{\omega}}\,{\cal S}_n^+.
\eeq
${\cal S}_n^-$ is not analytic in $E$ but has simple poles at
\beq
E=n\omega-\delta-\frac{g^2}{\omega},~~~~n=0,1,\cdots.
   \label{Rabi-exceptional2}
\eeq
which gives the other set of exact energies of the driven Rabi model.

The energies (\ref{Rabi-exceptional1}) and (\ref{Rabi-exceptional2}) appear for special values of model parameters and 
correspond to the exceptional solutions of the driven Rabi model. 
When (\ref{Rabi-exceptional1}) and (\ref{Rabi-exceptional2}) are satisfied, the infinite series
expansions (\ref{Rabi-series-solution}) and (\ref{Rabi-series-solution2}) truncate and reduce to polynomials in $z$ but only 
if the system parameters satisfy certain constraints. This can be easily verified by following
the procedure in \cite{Zhang13b} (and thus the driven Rabi model is quasi-exactly solvable).  
Majority part of 
the spectrum of the driven Rabi model is regular for which (\ref{Rabi-exceptional1}) and (\ref{Rabi-exceptional2}) are not satisfied.
The regular spectrum of the model can be obtained as follows.

{}From (\ref{Rabi-diff2}) and (\ref{Rabi-diff1b}), we obtain the 3-term recurrence relation for ${\cal R}_n^-$ and ${\cal S}_n^+$, respectively,
\beqa
&&{\cal R}_1^-+X_0\,{\cal R}_0^-=0,\n
&&{\cal R}_{n+1}^-+X_n\,{\cal R}_n^-+Y_n\,{\cal R}_{n-1}^-=0,~~~n\geq 1,\label{Rabi-3term}
\eeqa
\beqa
&&{\cal S}_1^++U_0\,{\cal S}_0^+=0,\n
&&{\cal S}_{n+1}^++U_n\,{\cal S}_n^++V_n\,{\cal S}_{n-1}^+=0,~~~n\geq 1,\label{Rabi-3term2}
\eeqa
where
\beqa
&&X_n=\frac{1}{2g(n+1)}\left[E-n\omega+\delta-\frac{3g^2}{\omega}+\frac{\Delta^2}
   {E-n\omega-\delta+\frac{g^2}{\omega}}\right],\n
&&U_n=\frac{1}{2g(n+1)}\left[E-n\omega-\delta-\frac{3g^2}{\omega}+\frac{\Delta^2}
   {E-n\omega+\delta+\frac{g^2}{\omega}}\right],\n  
&&Y_n=V_n=\frac{1}{n+1}.
\eeqa
The characteristic equation for the $n\geq 1$ part of both (\ref{Rabi-3term}) 
and (\ref{Rabi-3term2}) is given by
$t^2-\frac{\omega}{2g}\,t=0$, which gives two distinct roots $t_1=0$ and $t_2=\frac{\omega}{2g}$.
 
Let ${\cal R}_{n, 1}^-$, ${\cal R}_{n, 2}^-$ and ${\cal S}_{n, 1}^+$, ${\cal S}_{n, 2}^+$ denote  the two linearly independent solutions of
the 2nd equation of (\ref{Rabi-3term}) and (\ref{Rabi-3term2}), respectively. Applying the Poincar\'e-Perron theorem (i.e. theorems 2.1 and 2.2 of \cite{Gautschi67}),  these solutions satisfy (${\cal Z}={\cal R}^-$ or ${\cal S^+}$)
\beq
\lim_{n\rightarrow\infty}\,\frac{{\cal Z}_{n+1, r}}{{\cal Z}_{n, r}}=t_r,~~~~~r=1, 2.
\eeq
Thus ${\cal R}_{n}^{min}\equiv {\cal R}_{n, 1}^-$ and ${\cal S}_{n}^{min}\equiv {\cal S}_{n, 1}^+$ are minimal solutions. 
It is not difficult to show that the infinite series expansions (\ref{Rabi-series-solution}) and (\ref{Rabi-series-solution2}) with
coefficients given respectively by the minimal solution ${\cal R}_{n}^{min}$ 
and ${\cal S}_{n}^{min}$ are entire functions, i.e the elements of the BH space.

We now proceed to find energy eigenvalues $E$ corresponding to the
minimal solutions ${\cal R}_n^{min}$ and ${\cal S}_n^{min}$.
We follow a procedure presented in \cite{Leaver86} that uses the relationship between minimal
solutions and infinite continued fractions \cite{Gautschi67}. A similar procedure was applied in 
\cite{Moroz12} to analyze the Rabi model.

The ratio of successive elements of the minimal solution sequences ${\cal R}_{n}^{min}$ and ${\cal S}_{n}^{min}$ 
are expressible in terms of continued fractions,
\beq
T_{n}=\frac{{\cal R}_{n+1}^{min}}{{\cal R}_n^{min}}=-\frac{Y_{n+1}}{~X_{n+1}-}\,\frac{Y_{n+2}}{~X_{n+2}-}\,
\frac{Y_{n+3}}{~X_{n+3}-}\,\cdots,   \label{Rabi-continued-fraction}
\eeq
\beq
T'_{n}=\frac{{\cal S}_{n+1}^{min}}{{\cal S}_n^{min}}=-\frac{V_{n+1}}{~U_{n+1}-}\,\frac{V_{n+2}}{~U_{n+2}-}\,
\frac{V_{n+3}}{~U_{n+3}-}\,\cdots,   \label{Rabi-continued-fraction2}
\eeq
which for $n=0$ reduce to, respectively
\beqa
T_{0}&=&\frac{{\cal R}_{1}^{min}}{{\cal R}_0^{min}}=-\frac{Y_{1}}{~X_{1}-}\,\frac{Y_{2}}{~X_{2}-}\,
\frac{Y_{3}}{~X_{3}-}\,\cdots.   \label{Rabi-continued-fraction-T0}\\
T'_{0}&=&\frac{{\cal S}_{1}^{min}}{{\cal S}_0^{min}}=-\frac{V_{1}}{~U_{1}-}\,\frac{V_{2}}{~U_{2}-}\,
\frac{V_{3}}{~U_{3}-}\,\cdots.   \label{Rabi-continued-fraction2-T0}
\eeqa
Note that the ratios $T_0=\frac{{\cal R}_1^{min}}{{\cal R}_0^{min}}$ and
$T'_0=\frac{{\cal S}_1^{min}}{{\cal S}_0^{min}}$ involve ${\cal R}_0^{min}$ and
${\cal S}_0^{min}$, although the above continued fraction expressions  are obtained from the 2nd equation of (\ref{Rabi-3term}) and (\ref{Rabi-3term2}) i.e those recurrence relations for $n\geq 1$. 
However, for the single-ended sequences appearing in the infinite series expansions
(\ref{Rabi-series-solution}) and (\ref{Rabi-series-solution2}),  the first equation (i.e. the $n=0$ part) of 
the recurrences (\ref{Rabi-3term}) and (\ref{Rabi-3term2}) requires that
\beqa
 T_0&=&-X_0=-\frac{1}{2g}\left(E+\delta-\frac{3g^2}{\omega}+\frac{\Delta^2}
    {E-\delta+\frac{g^2}{\omega}}\right).   \label{Rabi-continued-fraction-T01}\\
 T'_0&=&-U_0=-\frac{1}{2g}\left(E-\delta-\frac{3g^2}{\omega}+\frac{\Delta^2}
    {E+\delta+\frac{g^2}{\omega}}\right).   \label{Rabi-continued-fraction2-T01}\
\eeqa
In general, (\ref{Rabi-continued-fraction-T0}), (\ref{Rabi-continued-fraction-T01}), (\ref{Rabi-continued-fraction2-T0}) and (\ref{Rabi-continued-fraction2-T01}) can not be satisfied at the same time for arbitrary values of the recurrence coefficients $X_n$, $Y_n$, $U_n$ and $V_n$.  
Physical meaningful solutions are those that are element of the BH space
\cite{Schweber67}. They can be obtained if $E$
can be adjusted so that equations (\ref{Rabi-continued-fraction-T0}), (\ref{Rabi-continued-fraction-T01}), (\ref{Rabi-continued-fraction2-T0}) and (\ref{Rabi-continued-fraction2-T01}) are all satisfied. 
Equating the right hand sides of (\ref{Rabi-continued-fraction-T0}), (\ref{Rabi-continued-fraction2-T0}) and (\ref{Rabi-continued-fraction-T01}), 
(\ref{Rabi-continued-fraction2-T01}), respectively, we obtain two 
transcendental functions $Q(E)=T_0+X_0$ and $P(E)=T'_0+U_0$ for the spectrum $E$, where $T_0$, $T'_0$ are the continued fractions in (\ref{Rabi-continued-fraction-T0}) and (\ref{Rabi-continued-fraction2-T0}), while $X_0$, $U_0$ are 
the right hand sides of (\ref{Rabi-continued-fraction-T01}) and (\ref{Rabi-continued-fraction2-T01}). Then the
 zeros of $Q(E)$ and $P(E)$ correspond to the points in the parameter space where the conditions 
(\ref{Rabi-continued-fraction-T01}) and (\ref{Rabi-continued-fraction2-T01}) are satisfied, i.e. the regular energies of the driven Rabi model are given by the zeros of the transcendental functions.
The transcendental eigenvalue equations $Q(E)=0$ and $P(E)=0$  may be solved for $E$ by standard root-search techniques 
(see e.g. \cite{Leaver86,Liu92} and references therein).
 Only for the denumerable infinite values of $E$ which are the roots of $Q(E)=0$ and $P(E)=0$, do
we get entire solutions of the differential equations (\ref{Rabi-diff1}), (\ref{Rabi-diff2}), (\ref{Rabi-diff1b}) and (\ref{Rabi-diff2b}).

\section{2-photon quantum Rabi model}\label{2-photon}
The Hamiltonian of the 2-photon Rabi model reads 
\begin{equation}
H_{2p}=\omega b^\dagger b+\Delta\sigma_z+g\,\sigma_x\left[(b^\dagger)^2 +b^2\right].\label{2-photon-RabiH}
\end{equation}
Let us make the canonical Bogoliubov transformation from $b, b^\dagger$ to squeezed bosons $a, a^\dagger$ \cite{Emary02},
\beq
b=\frac{a+\tau\, a^\dagger}{\sqrt{1-\tau^2}}, ~~~~~~b^\dagger=\frac{\tau\,a+a^\dagger}{\sqrt{1-\tau^2}},
\eeq
where $|\tau|<1$ is a real parameter. In terms of the squeezed bosons, the Hamiltonian (\ref{2-photon-RabiH})
takes the form
\beqa
\tilde{H}_{2p}&=&\Delta\sigma_z+\frac{1}{1-\tau^2}\left[\left(\omega\tau+g\sigma_x(1+\tau^2)\right)\left((a^\dagger)^2 +a^2\right)\right.\n
 & &\left.+\left(\omega(1+\tau^2)+4g\tau\sigma_x\right)a^\dagger a+\omega\tau^2+2g\tau\sigma_x\right].
 \label{2-photon-RabiH1}
\eeqa
Introduce the operators $K_\pm, K_0$ 
\beq
K_+=\frac{1}{2}(a^\dagger)^2,~~~~K_-=\frac{1}{2}a^2,~~~~K_0=\frac{1}{2}\left(a^\dagger a+\frac{1}{2}\right).
   \label{2-photon-Rabi-boson}
\eeq
Then (\ref{2-photon-RabiH1}) becomes
\beqa
\tilde{H}_{2p}&=&\Delta\sigma_z+\frac{1}{1-\tau^2}\left[2\left(\omega\tau+g\sigma_x(1+\tau^2)\right)(K_++K_-)\right.\n
& &\left.+2\left(\omega(1+\tau^2)+4g\tau\sigma_x\right)K_0\right]-\frac{1}{2}\omega.
\eeqa

The operators $K_\pm, K_0$ form the $su(1,1)$ Lie algebra. Its quadratic Casimir,
$C=K_+K_--K_0(K_0-1)$, takes the particular values $C=\frac{3}{16}$ in the representation 
(\ref{2-photon-Rabi-boson}).  This is the well-known infinite-dimensional unitary irreducible
representation ${\cal D}^+(q)$ of $su(1,1)$ with $q=\frac{1}{4},  \frac{3}{4}$.
Thus the Fock-Hilbert space decomposes into the direct sum of two subspaces ${\cal H}^q$ 
labeled by $q=1/4, 3/4$.

In the same way as the differential realization of boson operators in a Hilbert space of entire functions, we can 
represent \cite{Zhang13a,Zhang13b}  the generators $K_\pm, K_0$ (\ref{2-photon-Rabi-boson}) as single-variable differential operators in Bargmann space ${\cal B}_q$ with basis vectors given by the monomials $\left\{{z^n}/{\sqrt{[2(n+q-1/k^2)]!}}\right\}$, 
\beq
{K}_0 = z\frac{d}{dz}+ q, ~~~~{K}_+ = \frac{z}{2}, ~~~~~~
{K}_- =2z\frac{d^2}{dz^2}+4q\frac{d}{dz}.\label{2-photon-Rabi-diff}
\eeq
The Bargmann space ${\cal B}_q$ is the Hilbert space of entire functions on the complex plane if the inner product
\beq
(f,g)_q=\int\,\overline{f(z)}\,g(z)\,d\mu_q(z)\label{scalar-product1}
\eeq
is finite for an appropriate measure $d\mu_q(z)$. Taking the measure to be \cite{Zhang17}
\beq
d\mu_q(z)=\frac{1}{\pi}|z|^{2(q-3/4)}\,e^{-|z|}\,dx\,dy,
\eeq
then we can show by means of the formula $\Gamma(s)=\int^\infty_0\xi^{s-1}e^{-\xi}d\xi$ for ${\rm Re}(s)>0$,
\beq
(z^m, z^n)=\left[2\left(n+q-{1}/{4}\right)\right]!\,\delta_{m n}.\label{normalization1}
\eeq
Thus the monomials $\left\{{z^n}/{\sqrt{[2(n+q-1/k^2)]!}}\right\}$ form an orthonormal basis 
of the BH space ${\cal B}_q$ . Note in passing that the standard measure $\frac{1}{\pi}\,e^{-|z|^2}\,dx\,dy$ is
no longer appropriate here. It is now not difficult to see that if $f(z)=\sum_{n=0}^\infty\,
c_n z^n$ then
\beq
||f||_q^2=\sum_{n=0}^\infty \,|c_n|^2\,\left[2\left(n+q-{1}/{4}\right)\right]!\label{entireness-test1}
\eeq
and $f(z)$ is entire function in the BH space ${\cal B}_q$ if the sum on the right hand side converges. 

Using this differential realization, working in a representation defined by $\sigma_x$ diagonal
and choosing $\tau$ to be the root of $\omega\tau+g(1+\tau^2)=0$ so that it is real and obeys $|\tau|<1$, i.e.
\beq
\tau=-\frac{\omega}{2g}(1-\Omega),~~\Omega=\sqrt{1-\frac{4g^2}{\omega^2}},\label{tau}
\eeq
where $\left|\frac{2g}{\omega}\right|<1$, then the transformed 2-photon Rabi Hamiltonian becomes 
the matrix differential operator
\beq
\tilde{H}_{2p}=\left(
\begin{array}{cc}
2\omega\Omega\left(z\frac{d}{dz}+ q\right)-\frac{1}{2}\omega & \Delta  \\
\Delta & -\frac{8g}{\Omega}z\frac{d^2}{dz^2}+\frac{2}{\Omega}\left[\omega(2-\Omega^2)z-8gq\right]\frac{d}{dz} \\
 & -\frac{2g}{\Omega}z+\frac{2\omega(2-\Omega^2)q}{\Omega}-\frac{1}{2}\omega
\end{array}
\right). \label{2-photon-RabiH3}
\eeq

In terms of two-component wavefuntion $\psi(z)=\left(\psi_+(z), \psi_-(z)\right)^T$,
the time-independent Schr\"odinger equation, $\tilde{H}_{2p}\psi(z)=E\psi(z)$,
yields a system of coupled differential equations,
\beqa
&&\left[2\omega\Omega\left(z\frac{d}{dz}+ q\right)-\frac{1}{2}\omega-E\right]\psi_++\Delta\psi_-=0,
     \label{2-photon-diff1}\\
&&\left[8gz\frac{d^2}{dz^2}+\left(-2\omega(2-\Omega^2)z+16gq\right)\frac{d}{dz}   \right.\n
&&~~~~~~~~~\left. 
+2gz-2\omega(2-\Omega^2)q+(\frac{1}{2}\omega+E)\Omega\right]\psi_-  -\Omega\Delta\,\psi_+=0. 
     \label{2-photon-diff2}
\eeqa
This is a system of differential equations of Fuchsian type. 
Solutions to these equations must be analytic in the whole complex plane if
$E$ belongs to the spectrum of $\tilde{H}_{2p}$. So we are seeking solutions of the form
\beqa
\psi_+(z)=\sum_{n=0}^\infty\, {\cal K}_n^+(E)\, z^n, ~~~~
\psi_-(z)=\sum_{n=0}^\infty\, {\cal K}_n^-(E)\, z^n, \label{2-photon-series-solution}
\eeqa
which converge in the entire complex plane, i.e. solutions which are entire. 

Substituting (\ref{2-photon-series-solution}) into (\ref{2-photon-diff1}), we obtain
\beq
{\cal K}^+_n=\frac{\Delta}{E+\frac{1}{2}\omega-(2n+2q)\omega\Omega}{\cal K}_n^-.\label{2-photon-K+K-relation}
\eeq
So ${\cal K}^+_n$ is not analytic in $E$ but has simple poles at
\beq
E=-\frac{1}{2}\omega+(2n+2q)\omega\Omega,~~~~n=0, 1, \cdots.\label{2-photon-exceptional}
\eeq
The energies (\ref{2-photon-exceptional}) appear for special values of model parameters 
\cite{Emary02,Zhang13b} and 
correspond to the exceptional solutions of the 2-photon Rabi model. If (\ref{2-photon-exceptional})
is satisfied, the infinite series expansions
(\ref{2-photon-series-solution}) truncate and reduce to polynomials in $z$ but only if the system
parameters satisfy certain constraints \cite{Zhang13b}. 
Majority part of the spectrum of the 2-photon Rabi model is regular 
for which (\ref{2-photon-exceptional}) is not satisfied. 
The regular spectrum of the model is given by the zeros of the transcendental function $F(E)$ obtained below.
Thus similar to the Rabi case, the spectrum of the 2-photon Rabi model consists of two parts, the regular and
the exceptional spectrum.

{}From (\ref{2-photon-diff2}), we obtain the 3-step recurrence relation for ${\cal K}_n^-$ \cite{Zhang19},
\beqa
&&{\cal K}_1^-+A_0\,{\cal K}_0^-=0,\n
&&{\cal K}_{n+1}^-+A_n\,{\cal K}_n^-+B_n\,{\cal K}_{n-1}^-=0,~~~n\geq 1,   \label{2-photon-3-step}
\eeqa
where
\beqa
A_n&=&\frac{1}{8g(n+1)(n+2q)}\left[-(2n+2q)\omega(2-\Omega^2)\right.\n
& &     \left.+\left(E+\frac{1}{2}\omega-\frac{\Delta^2}{E+\frac{1}{2}\omega-(2n+2q)\omega\Omega}\right)\Omega\right],\n
B_n&=&\frac{1}{4(n+1)(n+2q)}.
\eeqa
The coefficients $A_n, B_n$ have the behavior as $n\rightarrow\infty$
\beq
A_n\sim a\,n^\alpha,~~~~~~B_n\sim b\,n^\beta
\eeq
with 
\beq
a=-\frac{\omega}{4g}(2-\Omega^2),~~~~\alpha=-1,~~~~b=\frac{1}{4},~~~~\beta=-2.\label{ab-values}
\eeq
Thus the asymptotic structure of solutions to the 2nd equation of (\ref{2-photon-3-step}) 
depends on the Newton-Puiseux
diagram formed with the points $P_0(0,0), P_1(1,-1), P_2(2,-2)$ \cite{Gautschi67}.
Let $\gamma$ be the slope of $\overline{P_0P_1}$ and $\delta$ the slope of $\overline{P_1P_2}$
so that $\gamma=\alpha$ and $\delta=\beta-\alpha$. Then we have $\gamma=\delta=\alpha$. The
characteristic equation of the $n\geq 1$ part of (\ref{2-photon-3-step}) reads 
$t^2+at+b=0$ with $a, b$ given in (\ref{ab-values}). It has two roots $t_1=\frac{\omega}{4g}$,
$t_2=\frac{g}{\omega}$. Remembering the condition $\left|\frac{2g}{\omega}\right|<1$, we have $|t_2|<|t_1|$.
Applying the Perron-Kreuser theorem (i.e. Theorem 2.3 of \cite{Gautschi67}), 
we conclude that the two linearly independent solutions ${\cal K}_{n,1}^-$ and ${\cal K}_{n,2}^-$
of the $n\geq 1$ part (i.e. the truly 3-term part) of (\ref{2-photon-3-step}) satisfy
\beq
\lim_{n\rightarrow\infty}\frac{{\cal K}_{n+1,r}^-}{{\cal K}_{n,r}^-}\sim t_r\,n^{-1},~~~~r=1,2.
   \label{K-asymptotics}
\eeq
So ${\cal K}_{n,2}^-$ is a minimal solution and ${\cal K}_{n,1}^-$ is a dominant one.
By (\ref{entireness-test1}), we can see that the infinite power series in 
(\ref{2-photon-series-solution}) with expansion coefficients ${\cal K}_{n,r}^-$ is entire if the sum
\beq
\sum_{n=0}^\infty \left|{\cal K}_{n,r}^-\right|^2[2(n+q-1/4)]!\label{K-sum}
\eeq
converges. Using the asymptotic form (\ref{K-asymptotics}) we get
\beq
\lim_{n\rightarrow\infty}\frac{\left|{\cal K}_{n+1,r}^-\right|^2[2(n+1+q-1/4)]!}
{\left|{\cal K}_{n,r}^-\right|^2[2(n+q-1/4)]!}=4\,|t_r|^2
\eeq
which is less than 1 for $r=2$ and greater than 1 for $r=1$. Thus by the ratio test, the sum
(\ref{K-sum}) converges for the minimal solution ${\cal K}_n^{min}\equiv {\cal K}_{n,2}^-$
and diverges for the dominant solution ${\cal K}_{n,1}^-$. It follows that
the infinite power series expansions $\psi_\pm^{min}(z)$, obtained
by substituting ${\cal K}_n^{min}$ for the ${\cal K}^-_n$'s in (\ref{2-photon-K+K-relation}) and 
(\ref{2-photon-series-solution}), converge in the whole complex plane, i.e. they are entire.

By the Pincherle theorem (i.e. Theorem 1.1 of \cite{Gautschi67}), the ratio of successive elements of
the minimal solution sequence ${\cal K}_n^{min}$ is expressible as continued fractions,
\beq
R_{n}=\frac{{\cal K}_{n+1}^{min}}{{\cal K}_n^{min}}=-\frac{B_{n+1}}{~A_{n+1}-}\,\frac{B_{n+2}}{~A_{n+2}-}\,
\frac{B_{n+3}}{~A_{n+3}-}\,\cdots,   \label{2-photon-continued-fraction}
\eeq
which for $n=0$ gives
\beq
R_{0}=\frac{{\cal K}_{1}^{min}}{{\cal K}_0^{min}}=-\frac{B_{1}}{~A_{1}-}\,\frac{B_{2}}{~A_{2}-}\,
\frac{B_{3}}{~A_{3}-}\,\cdots.   \label{2-photon-continued-fraction1}
\eeq
Note that the ratio $R_0=\frac{{\cal K}_1^{min}}{{\cal K}_0^{min}}$ involves ${\cal K}_0^{min}$, 
although the above continued fraction expression is obtained from the 
2nd equation of (\ref{2-photon-3-step}), i.e the recurrence (\ref{2-photon-3-step}) for $n\geq 1$. 
On the other hand, the ratio $R_0=\frac{{\cal K}_1^{min}}{{\cal K}_0^{min}}$ of 
the first two terms of a minimal solution is unambiguously fixed by the first equation of 
the recurrence (\ref{2-photon-3-step}), namely,
\beq
 R_0=-A_0=\frac{1}{16gq}\left[2q\omega(2-\Omega^2)
   -\left(E+\frac{1}{2}\omega-\frac{\Delta^2}{E+\frac{1}{2}\omega-2q\omega\Omega}\right)\Omega\right].   
   \label{2-photon-continued-fraction2}
\eeq
In general, the $R_0$ computed from the continued fraction (\ref{2-photon-continued-fraction1}) 
can not be the same as that from (\ref{2-photon-continued-fraction2})
for arbitrary values of
recurrence coefficients $A_n$ and $B_n$. Following similar discussions in last section for regular energy spectrum of the driven Rabi model, entire analytic function solutions which are elements of the BH space ${\cal B}_q$ require $E$ be the roots of the
transcendental equation 
%
$F(E)=R_0+A_0=0$ with $R_0$ given by the continued fraction in (\ref{2-photon-continued-fraction1}).   In other words, $F(E)=0$ is the eigenvalue equation of the 2-photon Rabi model.
 Only for the denumerable infinite values of $E$ which are the roots of $F(E)=0$, do
we get entire analytic function solutions of the differential equations (\ref{2-photon-diff1}) and (\ref{2-photon-diff2}) which are normalizable with respect to the BH norm (\ref{scalar-product1}).

\section{Two-mode quantum Rabi model}\label{two-mode}
We consider the Hamiltonian of the two-mode Rabi model introduced in \cite{Zhang13b}
\beq
H_{2m}=\omega(b_1^\dagger b_1+b_2^\dagger b_2)+\Delta\sigma_z+g\,\sigma_x(b_1^\dagger b_2^\dagger +b_1 b_2),
  \label{2-mode-RabiH}
\eeq
where we assume that the boson modes are degenerate with the same frequency $\omega$.
Introduce the two-mode Bogoliubov transformation,
\beqa
b_1=\frac{a_1+\sigma\, a_2^\dagger}{\sqrt{1-\sigma^2}},~~~~
  b_1^\dagger=\frac{\sigma\, a_2+a_1^\dagger}{\sqrt{1-\sigma^2}},~~~~ 
b_2=\frac{a_2+\sigma\, a_1^\dagger}{\sqrt{1-\sigma^2}},~~~~
  b_2^\dagger=\frac{\sigma\, a_1+a_2^\dagger}{\sqrt{1-\sigma^2}}.\label{bogoliubov-for-2-mode}
\eeqa
Here $|\sigma|<1$ is a real parameter and $a_1, a_2, a_1^\dagger, a_2^\dagger$
are squeezed bosons satisfying the canonical commutation relations 
$[a_i, a_i^\dagger]=1,~~[a_i, a_j]=[a_i, a_j^\dagger]=[a_i^\dagger, a_j^\dagger]=0,~i,j=1,2$.
In terms of the 2-mode squeezed bosons, the Hamiltonian (\ref{2-mode-RabiH}) has the form
\beqa
\tilde{H}_{2m}&=&\frac{1}{{1-\sigma^2}}\left[\left(2\omega\sigma+g\sigma_x(1+\sigma^2)\right)
    (a_1^\dagger a_2^\dagger+a_1 a_2)\right.\n
& &+\left(\omega(1+\sigma^2)+2g\sigma\,\sigma_x\right)(a_1^\dagger a_1+a_2^\dagger a_2) 
\left. +2\omega\sigma^2+2g\sigma\,\sigma_x\right]+\Delta\sigma_z. \label{2-mode-RabiH1}
\eeqa
Introduce the operators $K_\pm, K_0$
\beq
K_+=a_1^\dagger a_2^\dagger,~~~~K_-=a_1 a_2,~~~~K_0=\frac{1}{2}(a_1^\dagger a_1+a_2^\dagger a_2+1).\label{2-mode-boson}
\eeq
Then (\ref{2-mode-RabiH1}) becomes
\beqa
\tilde{H}_{2m}&=&\Delta\sigma_z+\frac{1}{{1-\sigma^2}}\left[\left(2\omega\sigma+g\sigma_x(1+\sigma^2)\right)
    (K_++K_-)\right.\n
& &\left. +2\left(\omega(1+\sigma^2)+2g\sigma\sigma_x\right)K_0\right] -\omega.\label{2-mode-RabiH2}
\eeqa

The operators $K_\pm, K_0$ form the $su(1,1)$ Lie algebra. Its quadratic Casimir, $C=K_+K_--K_0(K_0-1)$, 
takes the particular values $C=\kappa(1-\kappa)$ in the representation (\ref{2-mode-boson}), 
where $\kappa=1/2, 1, 3/2,\cdots$. This is the well-known infinite-dimensional unitary 
irreducible representation of $su(1,1)$ known as the positive discrete series 
${\cal D}^+(\kappa)$. Thus the Fock-Hilbert space decomposes 
into the direct sum of infinite subspaces ${\cal H}^\kappa$ labeled by $\kappa=1/2, 1, 3/2, \cdots$.


Similar to the 2-photon Rabi case, we can represent \cite{Zhang13a,Zhang13b}  the generators $K_\pm, K_0$ (\ref{2-mode-boson}) 
as single-variable differential operators in $z$ in Bargmann space ${\cal B}_\kappa$ with
basis vectors given by the monomials $\left\{{z^n}/{\sqrt{n!(n+2\kappa-1)!}}\right\}$, 
\beq
K_0=z\frac{d}{dz}+\kappa,~~~~K_+=z,~~~~K_-=z\frac{d^2}{dz^2}+2\kappa\frac{d}{dz},
      \label{su11-diff-rep-2mode}
\eeq
where $\kappa=1/2, 1, 3/2,\cdots$.
The Bargmann space ${\cal B}_\kappa$ is the Hilbert space of entire functions in $z$ if the inner product
\beq
(f,g)_\kappa=\int\,\overline{f(z)}\,g(z)\,d\mu_\kappa(z)\label{scalar-product2}
\eeq
is finite for an appropriate measure $d_\kappa\mu(z)$. It can be shown \cite{Barut71} that if we choose
\beq
d\mu_\kappa(z)=\frac{4}{\pi}|z|^{2\kappa-1}\,K_{1/2-\kappa}(2|z|)\,dx\,dy,
\eeq
where $K_\nu(z)$ is the modified Bessel function of the third kind which has the Mellin transform
$\int^\infty_0\,2\xi^{\alpha+\beta}K_{\alpha-\beta}(2\xi^{1/2})\xi^{s-1}d\xi=\Gamma(s+2\alpha)\Gamma(s+2\beta)$,
then 
\beq
(z^m, z^n)_\kappa=n!\left(n+2\kappa-1\right)!\,\delta_{m n}.\label{normalization2}
\eeq
Thus the monomials $\left\{{z^n}/{\sqrt{n!(n+2\kappa-1)!}}\right\}$ form an orthonormal basis 
of the BH space ${\cal B}_\kappa$ . It is now not difficult to see that if $f(z)=\sum_{n=0}^\infty\,
c_n z^n$ then
\beq
||f||_\kappa^2=\sum_{n=0}^\infty \,|c_n|^2\,n!\left(n+2\kappa-1\right)!\label{entireness-test2}
\eeq
and $f(z)$ is entire function belonging to ${\cal B}_\kappa$ if the sum on the right hand side converges.

Using this differential realization, working in a representation defined by $\sigma_x$ diagonal
and choosing $\sigma$ to be the root of $2\omega\sigma+g(1+\sigma^2)=0$ so that it is real and satisfies $|\sigma|<1$, i.e. 
\beq
\sigma=-\frac{\omega}{g}(1-\Lambda),~~~~\Lambda=\sqrt{1-\frac{g^2}{\omega^2}},\label{sigma}
\eeq
where $\left|\frac{g}{\omega}\right|<1$, then the transformed  Hamiltonian (\ref{2-mode-RabiH2})
 becomes a matrix differential operator
\beq
\tilde{H}_{2m}=\left(
\begin{array}{cc}
2\omega\Lambda\left(z\frac{d}{dz}+\kappa\right)-\omega & \Delta \\
\Delta & -\frac{2g}{\Lambda}z\frac{d^2}{dz^2}+\frac{2}{\Lambda}\left[\omega(2-\Lambda^2)z-2g\kappa\right]\frac{d}{dz} \\
 & -\frac{2g}{\Lambda}z+\frac{2\omega(2-\Lambda^2)\kappa}{\Lambda}-\omega
\end{array}
\right).\label{2-mode-RabiH3}
\eeq
In terms of two-component wavefuntion $\phi(z)=\left(\phi_+(z), \phi_-(z)\right)^T$, 
the time-independent Schr\"odinger equation, $\tilde{H}_{2m}\phi(z)=E\phi(z)$,
yields a system of coupled differential equations,
\beqa
&&\left[2\omega\Lambda\left(z\frac{d}{dz}+\kappa\right)-\omega-E\right]\phi_++\Delta\phi_-=0,
   \label{2-mode-diff1}\\
&&\left[2gz\frac{d^2}{dz^2}+\left(-2\omega(2-\Lambda^2)z+4g\kappa\right)\frac{d}{dz}  \right.\n
&&~~~~~~~~~\left.
+2gz-2\omega(2-\Lambda^2)\kappa+(E+\omega)\Lambda\right]\phi_- -\Lambda\Delta\,\phi_+=0.
   \label{2-mode-diff2}
\eeqa
This is a system of differential equations of Fuchsian type. 
Solutions to these equations must be analytic in the whole complex plane if
$E$ belongs to the spectrum of $\tilde{H}_{2m}$. Similar to the 2-photon Rabi case, we seek solutions of the form
\beqa
\phi_+(z)=\sum_{n=0}^\infty\, {\cal Q}_n^+(E)\, z^n, ~~~~~~
\phi_-(z)=\sum_{n=0}^\infty\, {\cal Q}_n^-(E)\, z^n, \label{2-mode-series-solution}
\eeqa
which converge in the entire complex plane.

Substituting (\ref{2-mode-series-solution}) into (\ref{2-mode-diff1}), we obtain
\beq
{\cal Q}^+_n=\frac{\Delta}{E+\omega-(2n+2\kappa)\omega\Lambda}{\cal Q}_n^-.\label{2-mode-K+K-relation}
\eeq
So ${\cal Q}^+_n$ is not analytic in $E$ but has simple poles at
\beq
E=-\omega+(2n+2\kappa)\omega\Lambda,~~~~n=0, 1, \cdots.\label{2-mode-exceptional}
\eeq
The energies (\ref{2-mode-exceptional}) appear for special values of model parameters 
\cite{Zhang13b} and correspond to the
exceptional solutions of the two-mode Rabi model. If (\ref{2-mode-exceptional}) is satisfied,
the infinite series expansions (\ref{2-mode-series-solution}) truncate 
and reduce to polynomials in $z$ but only if the model parameters obey certain constraints \cite{Zhang13b}.
Majority part of the spectrum of the two-mode Rabi model is regular spectrum which does not have the form 
(\ref{2-mode-exceptional}). The regular spectrum of the model is given by the zeros of 
the transcendental function $G(E)$ obtained below. 
Thus again, the spectrum of the two-mode Rabi model consists of
two parts, the regular and the exceptional spectrum.

{}From (\ref{2-mode-diff2}), we obtain the 3-step recurrence relation for ${\cal Q}_n^-$
\cite{Zhang19},
\beqa
&&{\cal Q}_1^-+C_0\,{\cal Q}_0^-=0,\n
&&{\cal Q}_{n+1}^-+C_n\,{\cal Q}_n^-+D_n\,{\cal Q}_{n-1}^-=0,~~~n\geq 1, \label{2-mode-3-step}
\eeqa
where
\beqa
C_n&=&\frac{1}{2g(n+1)(n+2\kappa)}\left[-(2n+2\kappa)\omega(2-\Lambda^2)\right.\n
& &     \left.+\left(E+\omega-\frac{\Delta^2}{E+\omega-(2n+2\kappa)\omega\Lambda}\right)\Lambda\right],\n
D_n&=&\frac{1}{(n+1)(n+2\kappa)}.
\eeqa
The coefficients $C_n, D_n$ have the behavior as $n\rightarrow\infty$
\beq
C_n\sim c\,n^\mu,~~~~~~D_n\sim d\,n^\rho
\eeq
with 
\beq
c=-\frac{\omega}{g}(2-\Lambda^2),~~~~\mu=-1,~~~~d=1,~~~~\rho=-2.
\eeq
By analysis similar to the 2-photon case,  we see that the two linearly independent solutions 
${\cal Q}_{n,1}^-$ and ${\cal Q}_{n,2}^-$ of the $n\geq 1$ part
 of the recurrence (\ref{2-photon-3-step}) obey
\beq
\lim_{n\rightarrow\infty}\frac{{\cal Q}_{n+1,r}^-}{{\cal Q}_{n,r}^-}\sim t_r\,n^{-1},~~~~r=1,2,
\eeq
where $t_1=\frac{\omega}{g}, ~ t_2=\frac{g}{\omega}$ and $|t_2|<|t_1|$ (from the condition 
$\left|\frac{g}{\omega}\right|<1$). Thus ${\cal Q}_{n,2}^-$ is a minimal solution and 
${\cal Q}_{n,1}^-$ is a dominant one. Using (\ref{entireness-test2}) and by similar analysis to
the 2-photon case, we can conclude that 
the infinite power series expansions $\phi^{min}_\pm(z)$
generated by substituting the minimal solution ${\cal Q}_n^{min}\equiv {\cal Q}_{n,2}^-$ 
for the ${\cal Q}_n^-$'s in 
(\ref{2-mode-K+K-relation}) and (\ref{2-mode-series-solution}), converge in the whole complex plane. 

{}From the 2nd equation of (\ref{2-mode-3-step}), the ratio of successive elements
of the minimal solution ${\cal Q}_n^{min}$ can be expressed as continued fractions,
\beq
S_{n}=\frac{{\cal Q}_{n+1}^{min}}{{\cal Q}_n^{min}}=-\frac{D_{n+1}}{~C_{n+1}-}\,\frac{D_{n+2}}{~C_{n+2}-}\,
\frac{D_{n+3}}{~C_{n+3}-}\,\cdots,   \label{2-mode-continued-fraction}
\eeq
which for $n=0$ reduces to
\beq
S_{0}=\frac{{\cal Q}_{1}^{min}}{{\cal Q}_0^{min}}=-\frac{D_{1}}{~C_{1}-}\,\frac{D_{2}}{~C_{2}-}\,
\frac{D_{3}}{~C_{3}-}\,\cdots.   \label{2-mode-continued-fraction1}
\eeq
On the other hand, 
the ratio $S_0=\frac{{\cal Q}_1^{min}}{{\cal Q}_0^{min}}$ of the first two terms of 
a minimal solution is unambiguously fixed by  the $n=0$ part of 
the recurrence (\ref{2-mode-3-step}), that is,
\beq
S_0=-C_0=\frac{1}{4g\kappa}\left[2\kappa\omega(2-\Lambda^2)
   -\left(E+\omega-\frac{\Delta^2}{E+\omega-2\kappa\omega\Lambda}\right)\Lambda\right]   
   \label{2-mode-continued-fraction2}
\eeq
Then similar to the 2-photon Rabi case, entire power series solutions which are elements of the BH space ${\cal B}_\kappa$ require that 
$E$ can be adjusted so that equations (\ref{2-mode-continued-fraction1}) and (\ref{2-mode-continued-fraction2}) are both satisfied. 
This yields an implicit continued fraction equation for the regular spectrum $E$. That is, the regular energies $E$ of the two-mode Rabi model are determined by the zeros of the transcendental function $G(E)=S_0+C_0$ with $S_0$ and $C_0$ given by (\ref{2-mode-continued-fraction1}) and (\ref{2-mode-continued-fraction2}), respectively. Only for the denumerable infinite values of $E$ which are the roots of $G(E)=0$, do
we get entire analytic function solutions of the differential equations (\ref{2-mode-diff1}) and (\ref{2-mode-diff2}) which are normalizable with respect to the BH norm (\ref{scalar-product2}).

\section{Conclusions}\label{summary}
We presented the analytic representations of solutions to the driven, 2-photon and two-mode quantum Rabi models.
The regular eigenvalues of the models are given by the zeros of the transcendental 
functions, which have been analytically found by applying the Bogoliubov
transformations and the new BH spaces.  

This work should be of general interest due to its focus on the non-trivial generalizations of
the popular quantum Rabi model, but also be intriguing to those interested in analytic solutions 
of Fuchsian differential equations in physics.

\section*{Acknowledgments}
This work was partially supported by the Australian Research Council  Discovery-Projects grant DP190101529.

\bebb{99}

\bbit{Vedral06}
Vedral, V. (2006). Modern foundations of quantum optics. Imperial College Press, London.

\bbit{Khitrova06}
Khitrova, G.,  Gibbs, H.M.,  Kira, M., Koch, S.W.,  and Scherer, A. (2006). Vacuum Rabi splitting in semiconductor. {\em Nature Phys.}, {\bf 2}, 81.

\bbit{Thanopulos04}
Thanopulos, I.,  Paspalakis, E., and Kis, Z. (2004). Laser-driven coherent manipulation of molecular chirality. {\em Chem Phys. Lett.}, {\bf 390}, 228.

\bbit{Englund07}
Englund, D., Faraon, A., Fushman, I., Stoltz, N., Petroff, P., and Vuckovic, J. (2007). Controlling cavity reflectivity with a single quantum dot. {\em Nature}, {\bf 450}, 857.

\bbit{Niemczyk10}
Niemczyk, T., Deppe, F., Huebl, H., Menzel, E.P., Hocke, F., Schward, M.J., Garcia-Ripoll, J.J., Zueco, D., H\"ummer, T., Solano, E., Marx, A., and Gross, R. (2010). Circuit quantum elctrodynamics in ultrastrong-coupling regime. {\em Nature Phys.}, {\bf 6}, 772.

\bbit{Bargmann61}
Bargmann, V. (1961). On a Hilbert space of analytic functions and associated integral transform part I. {\em Comm. Pure Appl. Math.}, {\bf 14}, 187.

\bbit{Schweber67}
Schweber, S. (1967). On the application of Bargmann Hilbert spaces to dynamical problems. {\em Ann. Phys.}, {\bf 41}, 205.

\bbit{Braak11}
Braak, D. (2011). On the integrability of the Rabi model. {\em Phys. Rev. Lett.}, {\bf 107}, 100401.

\bbit{Solano11}
Solano, E. (2011). Viepoint: The dialogue between quantum light and matter. {\em Physics}, {\bf 4}, 68.

\bbit{Moroz12}
Moroz, A. (2012). On the spectrum of  class of quantum models. {\em Europhys. Lett.}, {\bf 100}, 60010. 

\bbit{Maciejewski12}
Maciejewski, A.J., Przybylska, M., and  Stachowiak, T. (2012). How to calculate spectra of Rabi and related models. arXiv:1210.1130 [math-ph].  

\bbit{Travenec12}
Trav\'enec, I. (2012). Solvability of the two-photon Rabi Hamiltonian. {\em Phys. Rev.}, A {\bf 85}, 043805.

\bbit{Chen12}
Chen, Q.H., Wang, C.,  He,  S., Liu, T., and Wang, K.L. (2012). Exact solvability of the quantum Rabi model using Bogoliubov operators. {\em Phys. Rev.}, A {\bf 86}, 023822.

\bbit{Moroz13}
Moroz, A. (2013). On solvability and integrability of the Rabi model. {\em Ann Phys.}, {\bf 338}, 319.

\bbit{Zhang13a}
Zhang, Y.-Z. (2013). Solving two-mode squeezed harmonic oscillator and $k$th-order harmonic generation in Bargmann-Hilbert spaces. {\em  J. Phys.}, A {\bf 46}, 455302.

\bbit{Zhang13b}
Zhang, Y.-Z. (2013). On the solvability of the quantum Rabi model and its 2-photon and two-mode generalizations. {\em J. Math. Phys.}, {\bf 54}, 102104.

\bbit{Zhong13}
Zhong, H.,  Xie,  Q.,  Batchelor, M.T., and Lee, C.  (2013). Analytical eigenstates for the quantum Rabi model. {\em J. Phys.},  A {\bf 46}, 415302.

\bbit{Moroz14}
Moroz, A. (2014). A hidden analytic structure of the Rabi model. {\em Ann Phys.}, {\bf 340}, 252.

\bbit{Zhang17}
Zhang, Y.-Z. (2017). On the 2-mode and $k$-photon quantum Rabi models. {\em Rev. Math. Phys.}, {\bf 29}, 1750013.

\bbit{Larson13}
Larson, J. (2013). Integrability vs quantum thermalization. {\em J. Phys.}, B {\bf 46}, 224016.

\bbit{Brune87}
Brune, M.,  Raimond, J.M.,  Goy, P.,  Davidovich,  L.,  and  Haroche, S. (1987).
Realization of a two-photon mass oscillator. {\em Phys. Rev. Lett.}, {\bf 59}, 1899.

\bbit{Valle10}
del Valle, E.,  Zippilli,  S., Laussy, F.P., Gonzalez-Tudela,  A.,  Morigi, G., and  Tejedor, C. (2012). Tow-photon lasing by a single quantum dot in a high-Q microcavity.
{\em Phys. Rev.}, B {\bf 81}, 035302.

\bbit{Ota11}
Ota, Y.,    Iwamoto, S.,   Kumagai, N., and Arakawa, Y. (2011). Spontaneous two photon emission from a single quantum dot. {\em Phys. Rev. Lett.}, {\bf 107}, 233602.

\bbit{Leaver86}
Leaver, E.W. (1986). Solutions to a generalized spheroidal wave equation: Teukolsky's equations in general relativity, and the two-center problem in molecular quantum mechanics. {\em J. Math. Phys.}, {\bf 27}, 1238.

\bbit{Gautschi67}
Gautschi, W. (1967). Computational aspects of three-term recurrence relations. {\em  SIAM Review}, {\bf 9}, 24.

\bbit{Emary02}
Emary, C.,  and Bishop, R.F. (2002). Exact isolated solutions for the two-photon Rabi Hamiltonian. {\em  J. Phys.},  A {\bf 35}, 8231.

\bibitem{Zhang19}
Zhang, Y.-Z. (2019). Quasi-exactly solvable models. Lectures given at the 9th NSFC Summer School in Theoretical Physics, Xi'an, 2016. In ``Integrable Models and Their Applications", edited by W.-L. Yang, Z.-Y. Yang and T. Yang, Science Press, China.

\bbit{Liu92}
Liu, J.W. (1992). Analytical solutions to the generalized spheroidal wave equation and the Green's function of one-electron diatomic molecules. {\em J. Math. Phys.}, {\bf 33}, 4026.

\bbit{Barut71}
Barut, A.O., and  Girardello, L. (1971). New ``coherent" states associated with non-compact groups. {\em Comm. Math. Phys.},  {\bf 21}, 41.

\eebb

\end{document}